\documentclass[twocolumn]{aastex631}

\usepackage{amsmath} 
\usepackage{color}

\begin{document}

\title{SN 2023vbg: A Type IIn Supernova Resembling SN 2009ip, with a Long-Duration Precursor and Early-Time Bump}

\author{Sota Goto}
\affiliation{Graduate School of Science and Engineering, Kagoshima University, 1-21-35 Korimoto,
Kagoshima, Kagoshima 890-0065, Japan}
\author{Masayuki Yamanaka}
\affiliation{Amanogawa Galaxy Astronomy Research Center (AGARC), Graduate School of Science and Engineering, Kagoshima University, 1-21-35 Korimoto, Kagoshima, Kagoshima
890-0065, Japan}
\author{Takahiro Nagayama}
\affiliation{Graduate School of Science and Engineering, Kagoshima University, 1-21-35 Korimoto,
Kagoshima, Kagoshima 890-0065, Japan}

\author{Keiichi Maeda}
\affiliation{Department ofAstronomy, Kyoto University, Kitashirakawa-Oiwake-cho, Sakyo-ku, Kyoto 606-8502, Japan}
\author{Miho Kawabata}
\affiliation{Department ofAstronomy, Kyoto University, Kitashirakawa-Oiwake-cho, Sakyo-ku, Kyoto 606-8502, Japan}

\author{D.K.Sahu}
\affiliation{Indian Institute of Astrophysics, Koramangala 2nd Block, Bangalore 560034, India}

\author{Avinash Singh}
\affiliation{Oskar Klein Centre, Department ofAstronomy, Stockholm University, AlbaNova, SE-106 91 Stockholm, Sweden}
\author{Anjasha Gangopadhyay}
\affiliation{Oskar Klein Centre, Department ofAstronomy, Stockholm University, AlbaNova, SE-106 91 Stockholm, Sweden}

\author{Naveen Dukiya}
\affiliation{Aryabhatta Research Institute of observational sciencES, Manora Peak, Nainital 263 001, India}
\author{Kuntal Misra}
\affiliation{Aryabhatta Research Institute of observational sciencES, Manora Peak, Nainital 263 001, India}
\author{Monalisa Dubey}
\affiliation{Aryabhatta Research Institute of observational sciencES, Manora Peak, Nainital 263 001, India}
\author{Bhavya Ailawadhi}
\affiliation{Aryabhatta Research Institute of observational sciencES, Manora Peak, Nainital 263 001, India}

\begin{abstract}
Type IIn supernovae (SNe) resembling SN 2009ip (09ip-like SNe) originate from the interaction between circumstellar material (CSM) and the ejecta.
This subclass not only shares similar observational properties around the maximum, but is commonly characterized by a long-duration precursor before its maximum. 
Investigating the observed properties of the precursor provides constraints on the mass-loss history of the progenitor.
We present observational data of SN 2023vbg, a 09ip-like type IIn SN that displayed unique observational properties compared to other 09ip-like SNe. SN 2023vbg showed a long-duration precursor at $M_g\sim-14$ mag lasting for $\sim100$ days, followed by a bright bump at $M_g\sim-17$ mag at 12-25 days before the maximum.
The luminosity of the precursor is similar to those of other 09ip-like SNe, but the bright bump has not been observed in other cases.
After reaching the peak luminosity, the light curve exhibited a relative smooth decline.
While the H$\alpha$ profile displays two velocity components ($\sim 500$ and $3000\ \mathrm{km\ s^{-1}}$), a broad component observed in other 09ip-like SNe was not seen, but it may emerge later. We suggest that these properties are explained by the difference in the CSM structure as compared to other 09ip-like SNe; SN 2023vbg had an inner denser CSM component, as well as generally smooth CSM density distribution in a more extended scale, than in the others. Such diversity of CSM likely reflects the diversity of pre-SN outbursts, which in turn may mirror the range of evolutionary pathways in the final stages of the progenitors.
\end{abstract}


\section{Introduction} \label{sec:intro}
Type IIn SNe are a subclass of Type II SNe, characterized by narrow Balmer series emission lines and a blue continuum \citep{schlegel1990,filippenko1997}.
The primary source of radiative energy is believed to be the interaction between the supernova (SN) ejecta 
and pre-existing CSM \citep[see][for review]{gal-yam2017,smith2017hos}.
Therefore, Type IIn SNe exhibit a higher degree of observational diversity compared to other types of SNe \citep[see][for sample studies]{taddia2013,Nnyholm2020,hiramatsu2024}.
This diversity may reflect differences in the progenitor's mass-loss history and the origin of CSM.
It is important to note that not all Type IIn supernovae necessarily share the same type of progenitor. \citep[e.g., 2022mop;][]{brennan2025}

Some Type IIn SNe exhibit remarkably pre-SN activity. A remarkable example is SN 2009ip \citep[e.g.,][]{mauerhan2013,pastorello2013,margutti2014}, and a group of objects that exhibit observational characteristics similar to SN 2009ip are referred to as `09ip-like' objects \citep[e.g.,][]{ofek2013,elias-rosa2016,pastorello2018,Fransson2022,pastorello2025}. 
They share very similar light curve properties, including the rise and decline timescales and peak luminosity, as well as spectral features \citep[see e.g.,][for a review]{fraser2013,fraser2020}. 

The most distinctive feature of 09ip-like objects is the luminosity variability observed months to years before the peak brightness, known as precursor events.
These precursor events are believed to reflect a variable mass-loss history of the progenitor star \citep[e.g.,][]{smith2014review}. A popular scenario suggests that the progenitor was a luminous blue variable (LBV) star undergoing a terminal explosive event \citep[e.g.,][]{mauerhan2013,margutti2014}.
However, some studies have proposed that these events are not necessarily terminal explosions, and the debate on their true nature remains ongoing \citep[e.g.,][]{thone2017,jencson2022}. 
Therefore, further observational studies are required to clarify the nature of 09ip-like objects.

Furthermore, the luminosity variations observed as precursor events are believed to be associated with ejection of matrials and formation of CSM. Placing observational constraints on the properties of these precursors is an important step toward identifying the progenitors of Type IIn SNe, especially in the case of 09ip-like events.

In this paper, we report on photometric and spectroscopic observations of SN 2023vbg, a peculiar Type IIn SN, from the ultraviolet (UV) to the near-infrared (NIR) wavelengths. 
This SN exhibited a two-stage precursor lasting approximately 100 days. 
Although SN 2023vbg displayed characteristics commonly seen in the 09ip-like SN class, it was found to possess distinct characteristics that highlight diverse properties among the class.

The structure of this paper is as follows.
In Sect.~\ref{sec:observation and data}, we describe the observations and the data acquisition process.
Sect.~\ref{sec:results} presents the observational results, followed by discussion in Sect.~\ref{sec:disccusion} on the possible nature of this SN based on these findings.
Sect.~\ref{sec:summary and conclusion} provides a summary of our main findings.

\section{Observation and data} \label{sec:observation and data}

\subsection{First Detection and Classification} \label{ssec:detection}
SN 2023vbg (also known as PS22nhw, ATLAS23yal and ZTF23abkhwgb) was first discovered by the Zwicky Transient Facility (ZTF) survey \citep{ZTF_Bellm,Graham2019}, using the Palomar Schmidt 48-inch (P48) Samuel Oschin Telescope on 2023 October 15.5 (MJD = 60232.487; $t=-97.8$ d\footnote{$t=0$ corresponds to the peak of the ATLAS $o$-band light curve (MJD=60330.36)}).
The first detection was in the $r$-band, with a difference magnitude of 20.4 $\pm$ 0.2 mag at coordinates $\alpha$ = 07h43m43s.812, $\delta$ = +34$^{\circ}$22$^{\prime}$29$^{\prime\prime}$.96 (J2000.0) \citep{ZTF_detection}.
A non-detection was recorded in the $g$-band at a limiting magnitude of 20.6 mag on 2023 October 15, one hour before the discovery. 
Subsequently, there was a gap of about 80 days in the evolution.
On December 30 ($t=-21.8$ d), a rapid rising that was found by Asteroid Terrestrial-impact Last Alert System \citep[ATLAS;][]{tonry2018} was reported by \citet{2024TNSAN..22....1P}, with an observed magnitude of 17.9 $\pm$ 0.2 mag in the orange-ATLAS filter.

SN 2023vbg was classified as a Type IIn SN at a redshift of $z = 0.0173$ by \citet{2024TNSCR..82....1W}, using a spectrum obtained with the Lick-3m KAST spectrograph on 2024 January 8 ($t=-12.8$ d from the SN peak).
Since the redshift of the host galaxy has not been published, this redshift value was determined from the H$\alpha$ line in the first obtained spectrum.
In this study, we adopt cosmological constants of $H_0 = 71\ \mathrm{km\,s^{-1}\,Mpc^{-1}}$, $\Omega_m = 0.3$, and $\Omega_\Lambda = 0.7$.
Based on these parameters, the calculated luminosity distance (calculated by {\sc cosmotool}
\footnote{\url{http://www.bo.astro.it/~cappi/cosmotools}}) 
is $ d_L = 72.7\ \mathrm{Mpc^{-1}}$, corresponding to a distance modulus of $\mu = 34.30$ mag.

\subsection{Photometry} \label{ssec:photo}
The photometric data presented in this paper is a combination of public data and those obtained by our observations. 
For the ZTF $gr$-bands, we used publicly-available data through the ALeRCE explorer\footnote{\url{https://alerce.science/}} \citep{sanchez2021,carrasco2021,foster2021}.
In addition, we utilized the ZTF forced-photometry service\footnote{\url{https://ztfweb.ipac.caltech.edu/cgi-bin/requestForcedPhotometry.cgi}} \citep{Masci2018} to perform forced photometry on archival images over the period from $t=-2135.8$ d to $t=332.6$ d, including a search for precursor events. For the ZTF forced-photometry data, detections and non-detections were determined based on the ratio of flux to its error (1$\sigma$ uncertainty). 
We defined cases with flux/err $\geq$ 5 as detections and those with flux/err $<$ 5 as non-detections. 
For photometric values classified as detections, magnitudes were calculated using $-2.5log_{10}(flux)+ZP$, where ZP is the zero point in the AB magnitude system.
For non-detections, upper limits for each epoch were adopted as $-2.5log_{10}(5 \times err)+ZP$.


UV and $UBV$-band photometric data were obtained from publicly available data from the UVOT at the Neil Gehrels Swift Observatory \citep{roming2005}.
Aperture photometry was performed using the UVOT data analysis software in HEASOFT \footnote{\url{https://heasarc.gsfc.nasa.gov/docs/software/heasoft/}}, following a standard procedure.
$UBV$-band data also were obtained from the 1.3-m Devasthal Fast Optical Telescope (DFOT) at the Aryabhatta Research Institute of Observational Sciences (ARIES).
Additionally, $gi$-band data obtained with the 1-m telescope at the Iriki Observatory using the dual-band optical imager were included in the dataset.
The analysis of these data from the DFOT and the Iriki 1-m telescope was performed using standard tasks included in the data processing software IRAF\footnote{IRAF stands for Image Reduction and Analysis Facility distributed by the National Optical Astronomy Observatory, operated by the Association of Universities for Research in Astronomy (AURA) under a cooperative agreement with the National Science Foundation.}/DAOPHOT \citep{stetson1987}. 
Primary processing and Point spread function (PSF) photometry were carried out as part of this analysis.

NIR data in the $JHKs$ bands were obtained using the kSIRIUS \citep{nagayama2024} which is mounted on the 1-m telescope at the Iriki Observatory.
These data were reduced using a standard procedure using IRAF, following a similar approach as for the optical data. PSF photometry was subsequently performed on the reduced NIR imaging data.

After $t=40$ d, contamination from the host galaxy becomes non-negligible in UV and NIR bands. Therefore, for UV and NIR bands, as well as some optical bands ($i$-band), only data before $t=40$ d were used in the subsequent analysis.

In our analysis, we corrected all photometric measurements for the Galactic extinction at the location of SN 2023vbg using a reddening value of  $E(B-V)_{MW} =0.05$ mag \citep{Finkbener2011}. 
For UV-bands, the method of \citet{Siegel2014} was applied to estimate the extinction for each band.
For optical and NIR extinction corrections, we applied the extinction law of \citet{Cardelli1989} with $R_V$ = 3.1.
As for the host galaxy extinction, no signs of Na {\sc i} D absorption were detected in any of our spectra (as shown in Sect.\ref{sec:results}). Therefore, no additional correction was applied.
We adopt these values throughout this paper.

\subsection{Optical Spectroscopy} \label{ssec:optical spectroscopy}
Optical spectroscopic observations were conducted using several telescopes.
Two epochs of spectra were obtained with Aries Devasthal Faint Object Spectrograph and Camera (ADFOSC) mounted on the 3.6m Devasthal Optical Telescope (DOT) at ARIES, on 2024 February 16 and March 8 ($t=26$ and $47$ d).
Additionally, two epochs of spectra were obtained on 2024 March 4 and March 24 ($t=43$ and $63$ d) using the Hanle Faint Object Spectrograph Camera (HFOSC) mounted on the 2-m Himalayan Chandra Telescope (HCT) of the Indian Institute of Astrophysics (IIA).
Furthermore, a spectrum was acquired on 2024 March 15 ($t=54$ d), using Kyoto Okayama Optical Low-dispersion Spectrograph with
optical-fiber Integral Field Unit \citep[KOOLS-IFU;][]{matsubayashi2019} mounted on the Kyoto University 3.8m Seimei Telescope.

We have also included the classification spectrum obtained with KAST, mounted on the 3m Lick Shane Reflector, which is publicly available on TNS \citep{2024TNSCR..82....1W}.

\section{Results} \label{sec:results}
\subsection{Photometric Evolution} \label{ssec:photometric evolution}

\begin{figure}[htbp]
    \centering
    \includegraphics[width=0.5\textwidth]{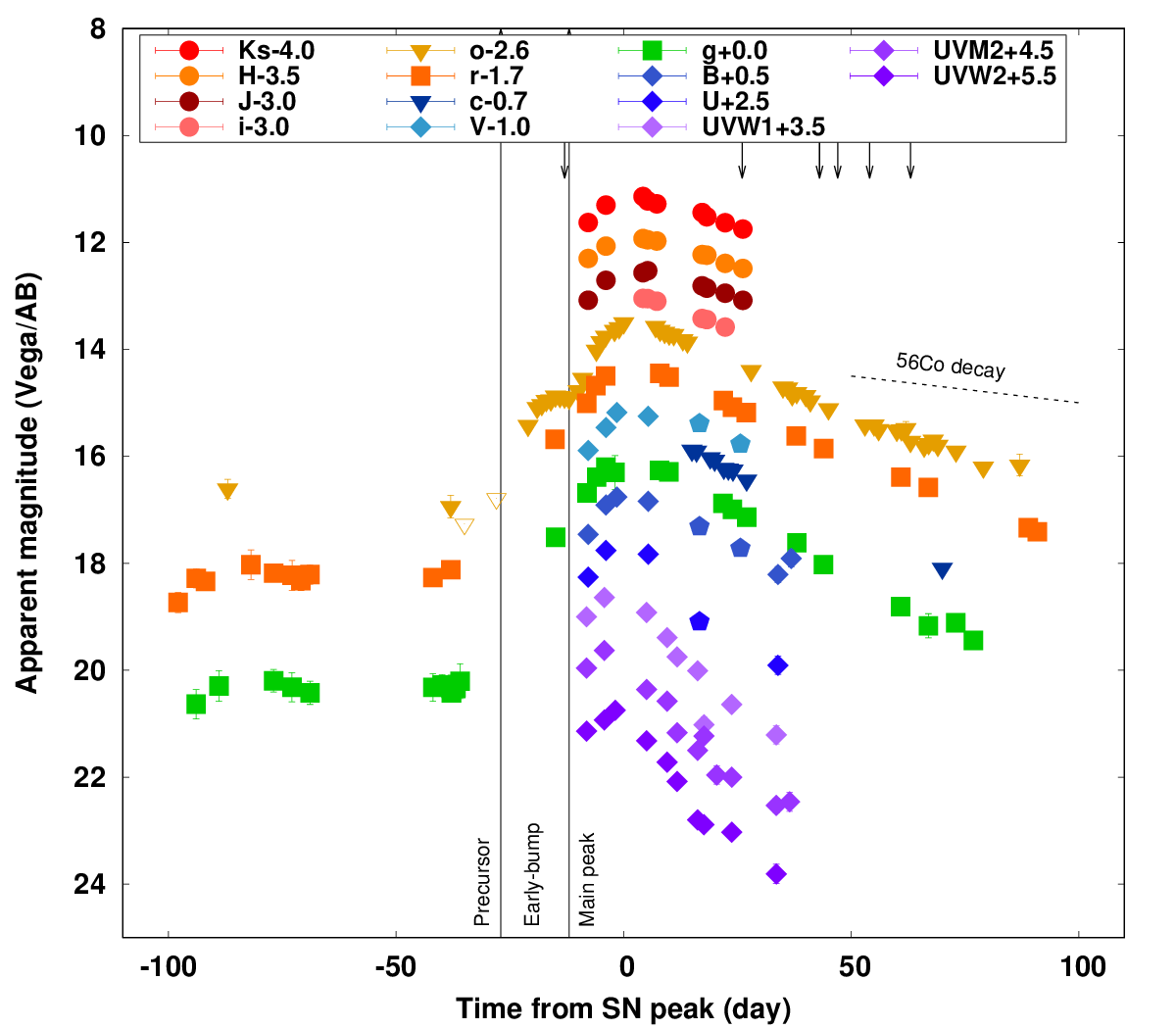}
    \caption{Multi-band light curves of SN 2023vbg. The vertical axis for $UBVJHKs$ bands is in Vega magnitudes, while the $gcro$ bands are in AB magnitudes. The circular data points represent data obtained at the Iriki Observatory, square data points correspond to ZTF, and downward triangles indicate ATLAS data and downward open triangles denote 3$\sigma$ upper limits from ATLAS. Diamonds represent UVOT data, while pentagons correspond to data from DFOT. The horizontal axis is referenced to the peak of the $o$-band. The black lines at the top of the figure indicate the dates of spectroscopic observations, and the black dotted lines mark the boundaries of each phase ($t=-27$ and $-12$ d, described in Sect.~\ref{ssec:photometric evolution}). Also, the decay rate of $^{56}$Co is shown as a black dotted line.}
    \label{fig1}
\end{figure}

The multi-band light curves of SN 2023vbg are shown in Figure~\ref{fig1}. As a notable feature, a precursor was observed before $t=-27$ d.
Subsequently, the light curve exhibits a flat evolution as found in the ATLAS observations.
Based on this light curve, we defined the phases as follows: the phase before $t=-27$ d is referred to as the `precursor', the phase from $t=-27$ d to $t=-12$ d as the `early-bump' and the phase after $t=-12$ d as the `Main peak'.
Subsequent analyses are based on these phase definitions.

Regarding the early-bump, unfortunately, it was observed only in the ATLAS $o$-band.
However, thanks to the high observational cadence of ATLAS, the shape of the bump is clearly captured.
Based on the light curve shape, we infer that the bump started to rise around $t=-27$ d.
Throughout this paper, we adopted the estimated explosion date as the start of the `early-bump' ($t=-27$ d) obtained by a fit using a quadratic function.
The `early-bump' ceased its rise by $t=-12$ d, transitioning into the next phase which is considered as the SN peak.
During the `Main peak' phase ($t>-12$ d), all bands reached their peak approximately 10 days after the `early-bump'. Following this, the optical and NIR bands exhibited a linear decline in luminosity, while UV bands showed a more rapid decline compared to the other bands.


\subsection{Precursor Activity} \label{ssec:precursor}
Between $t \sim-100$ and $-27$ d, the precursor activity was detected in the light curve obtained in the ZTF data through ALeRCE. 
To investigate whether any precursor activity could be identified prior to this detection, we analyzed data spanning approximately 2000 days before the SN light curve's peak, using the ZTF forced-photometry service.
The methods, including the criteria for detections and non-detections, are described in Sect.~\ref{ssec:photo}.

\begin{figure*}[htbp]
    \centering
    \includegraphics[width=0.8\textwidth]{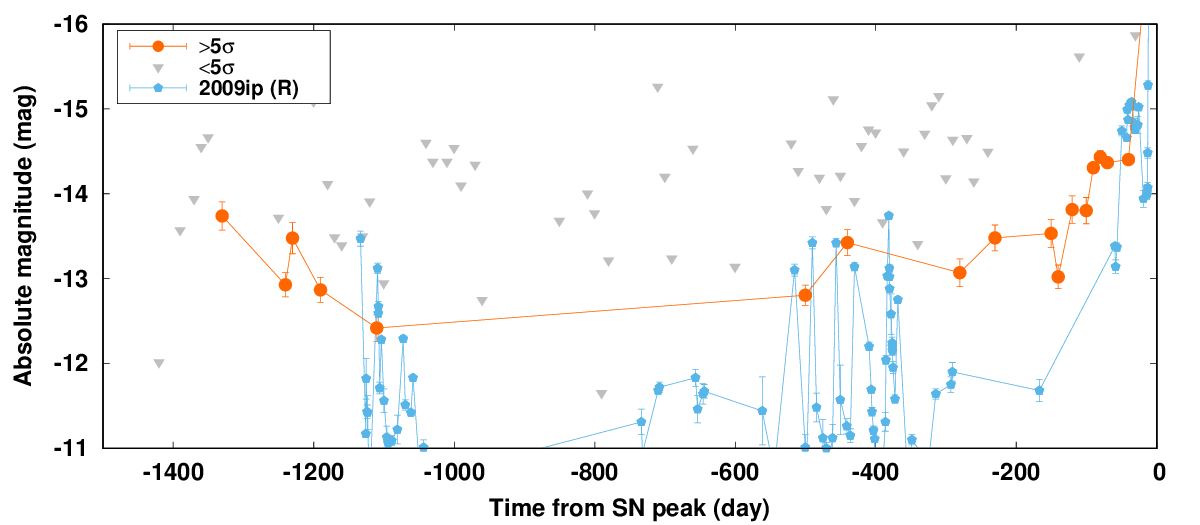}
    \caption{Pre-SN (precursor) phase $r$-band light curve of SN 2023vbg. For the comparison, we also plot the $R$-band light curve of SN 2009ip \citep{pastorello2013}. Observation up to 1500 days before the SN peak were obtained as a part of the ZTF survey. We binned the observational data at each epoch into 10-day intervals. The 5$\sigma$ upper limits are plotted as gray triangles, while detected observations are shown as filled circles.}
    \label{fig2}
\end{figure*}

In Figure~\ref{fig2}, the results of the ZTF forced-photometry ($r$-band), both 5$\sigma$ detections and upper limits for each epoch, are plotted. We found no 5$\sigma$ detection (using a 10-day binning) before $t\sim-1400$ d.
As shown in this figure, we detected some marginal precursor detections on the ZTF data.

The first marginal detection in the $r$-band occurred around $t\sim-1400$ d. 
Subsequently, multiple detections were observed between $t\sim-1400$ d and $-1100$ d, followed by another series of detections near $t\sim-400$ d. 
After $t\sim-200$ d, brighter detections - approximately 0.5 mag brighter than before - were recorded.
During this phase, the $r$-band brightened to $\sim-14.5$ mag and maintained this brightness as it transitioned into the `early-bump' phase after $t\sim-27$ d.
The detections roughly follow the lower edges of the non-detections, suggesting that the precursor persisted, even before $-200$ d, at the magnitude of $\sim-13$ mag for long time (more than 1000 days) with the variability of $\sim0.5-1$ mag, and it was detected only when the observation reached a sufficient depth depending on the observational conditions. 

\subsection{Light Curve Comparison}\label{ssec:LC compa}
\begin{figure}[htbp]
    \centering
    \includegraphics[width=0.5\textwidth]{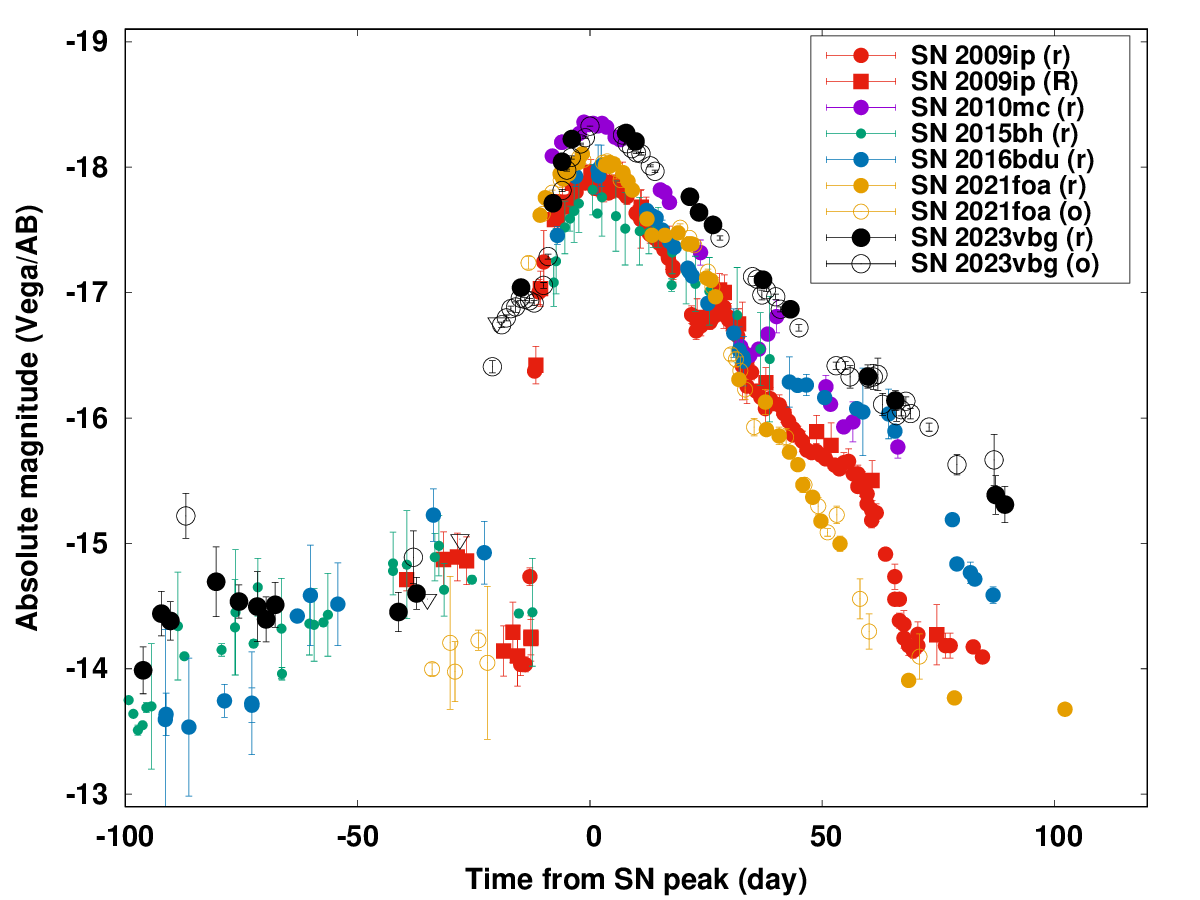}
    \caption{Comparison of $r/R/o$-band light curve of SN 2023vbg with SNe IIn 2009ip, 2010mc, 2015bh, 2016bdu and IIn/Ibn 2021foa. The $r/R$-band is represented by filled circles, while the $o$-band is shown as open circles and downward open triangles denote 3$\sigma$ upper limits. The data for the comparison objects have been corrected for extinction using the extinction values provided in each reference.}
    \label{fig3}
\end{figure}

Figure~\ref{fig3} compares the $r/R/o$-band light curves of SN 2023vbg and other 09ip-like objects from $t\sim-100$ d.
The comparison SNe IIn are as follows; SN 2009ip~\citep{fraser2013,mauerhan2013,graham2014}, SN 2010mc~\citep{ofek2013,smith2014}, SN 2015bh~\citep{elias-rosa2016}, SN 2016bdu~\citep{pastorello2018}, SN 2021foa~\citep{gangopadhyay2025}.
During the "precursor" phase ($t<-27$ d), SN 2023vbg exhibits luminosity variations between approximately $-14$ and $-15$ mag, a behavior similar to those of other objects. Additionally, the timescales of the rise and decline around the SN peak are quite similar to those of other 09ip-like SNe. 

However, two noteworthy points emerge from this comparison.
First, the `early-bump' observed between $t=-27$ and $t=-12$ d is unique to SN 2023vbg.
Notably, in SN 2009ip, the luminosity variations between the 2012a and 2012b events \citep[originally labeled as such by][]{pastorello2013} were observed with high cadence; during the phase of the `early-bump' in SN 2023vbg, a dimming of approximately 1 mag was observed in contrast to the rising bump seen in SN 2023vbg.
None of the other comparison objects exhibit a bump at the same phase as SN 2023vbg.

Second, the post-peak decline of SN 2023vbg follows a smooth decay, a feature that may be relatively uncommon in 09ip-like objects.
In contrast, SN 2009ip and other comparison objects display bumpy structures referred to as a "shoulder" \citep{gangopadhyay2025} or a "knee" \citep{reguitti2024} during the post-peak decline phase.
The monotonic decline observed in SN 2023vbg distinguishes it from the other objects.
These two features make SN 2023vbg a particularly intriguing object worthy of further attention.

\subsection{Temperature and Radius Evolution}\label{ssec:sed evolution}
\begin{figure}[htbp]
    \centering
    \includegraphics[width=0.5\textwidth]{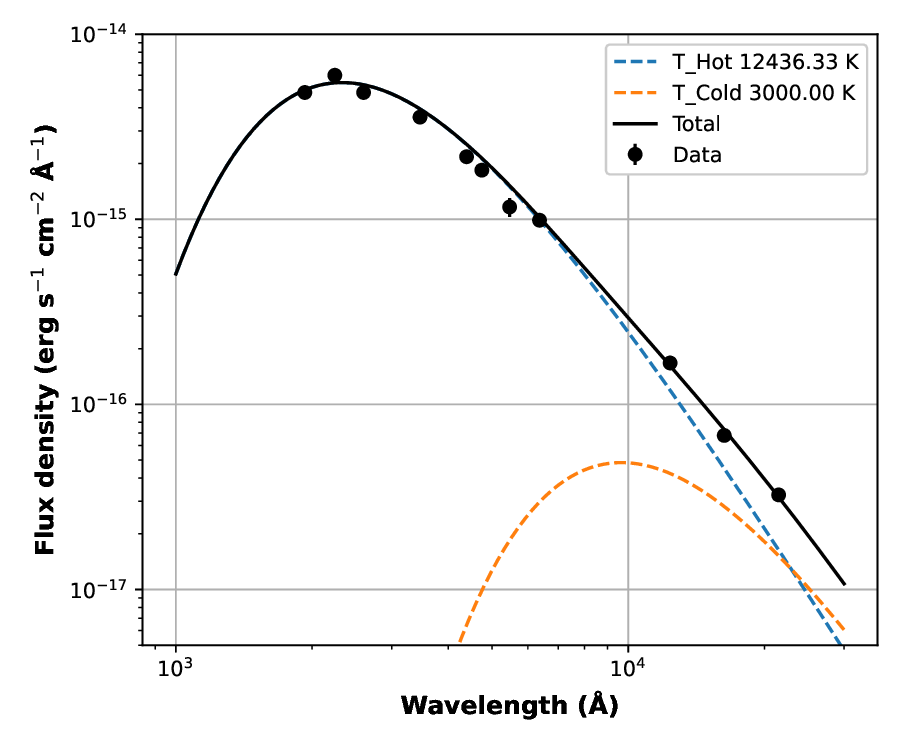}
    \caption{The BB fit results for the SED at $t=-3.8$~d, indicating an excess in NIR. The temperature of the `cold' component is fixed at $3000$~K.} 
    \label{fig4}
\end{figure}

\begin{figure}[htbp]
    \centering
    \includegraphics[width=0.5\textwidth]{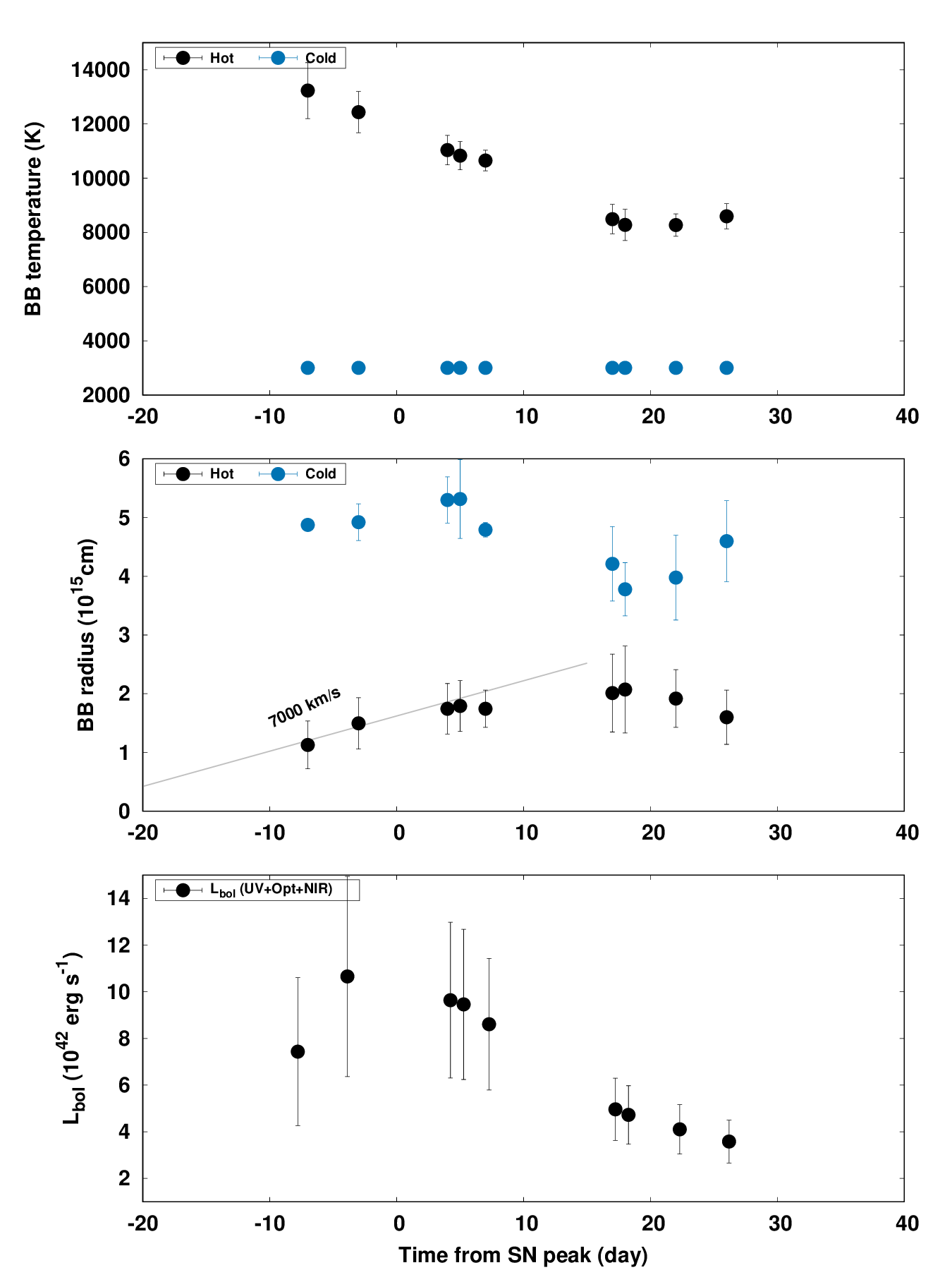}
    \caption{Upper panel: The evolution of the black body temperature estimated from the black body fitting. The black points represent the high-temperature component (`hot'), while the blue points represent the low-temperature component (`cold'). The temperature of cold component is fixed $3000$~K. Middle panel: The evolution of the effective radii estimated from the black body fitting. We draw line corresponding to velocities of $7000~\mathrm{km s^{-1}}$. Bottom panel: A UV-to-NIR bolometric light curve. We derived the luminosities by performing integration of the SED at each epoch.}
    \label{fig5}
\end{figure}

We investigated the time evolution of the spectral energy distribution (SED) of SN 2023vbg and performed fitting assuming a black body model. The SED was constructed by interpolating optical and ultraviolet data to match the epochs of NIR observations. The extinction correction for the UV flux has a relatively large uncertainty. Therefore, the black body fitting was performed using only wavelength data longer than 3000~\AA. 

Initially, a single black body fit was applied, but the early epochs showed a substantial NIR excess (Figure~\ref{fig4}).
To account for this excess, we performed a two-component black body fit. 
\cite{margutti2014} reported a similar early-time infrared excess in SN 2009ip.
To explain this excess, they performed a two-component black body fit and found that low-temperature component had a temperature of approximately 3000 K. 
In our fitting, we adopt this value (3000 K) as the temperature of the cool component.
Figure~\ref{fig5} presents the photospheric temperatures and the effective radii for the two components. 
Around the peak, the cold component was prominent, with an effective radius approximately five times larger than that of the hot component.

This result is consistent with the evolution of SN 2009ip reported by \cite{margutti2014}, although the origin of the cool component remains uncertain. One possible explanation is thermal emission from dust, as observed in some Type IIn supernovae. However, a temperature of 3000 K exceeds the typical dust sublimation temperature, suggesting that this component is too hot to be attributed to dust.

\cite{margutti2014} discussed the origin of this low-temperature component and concluded that it is likely due to free-free emission from ionized gas located in outer regions.
The low-temperature component observed in SN 2023vbg may have a similar origin.

\subsection{Spectroscopic Evolution} 
\label{ssec:spectra evolution}
\begin{figure*}[htbp]
    \centering
    \includegraphics[width=0.8\textwidth]{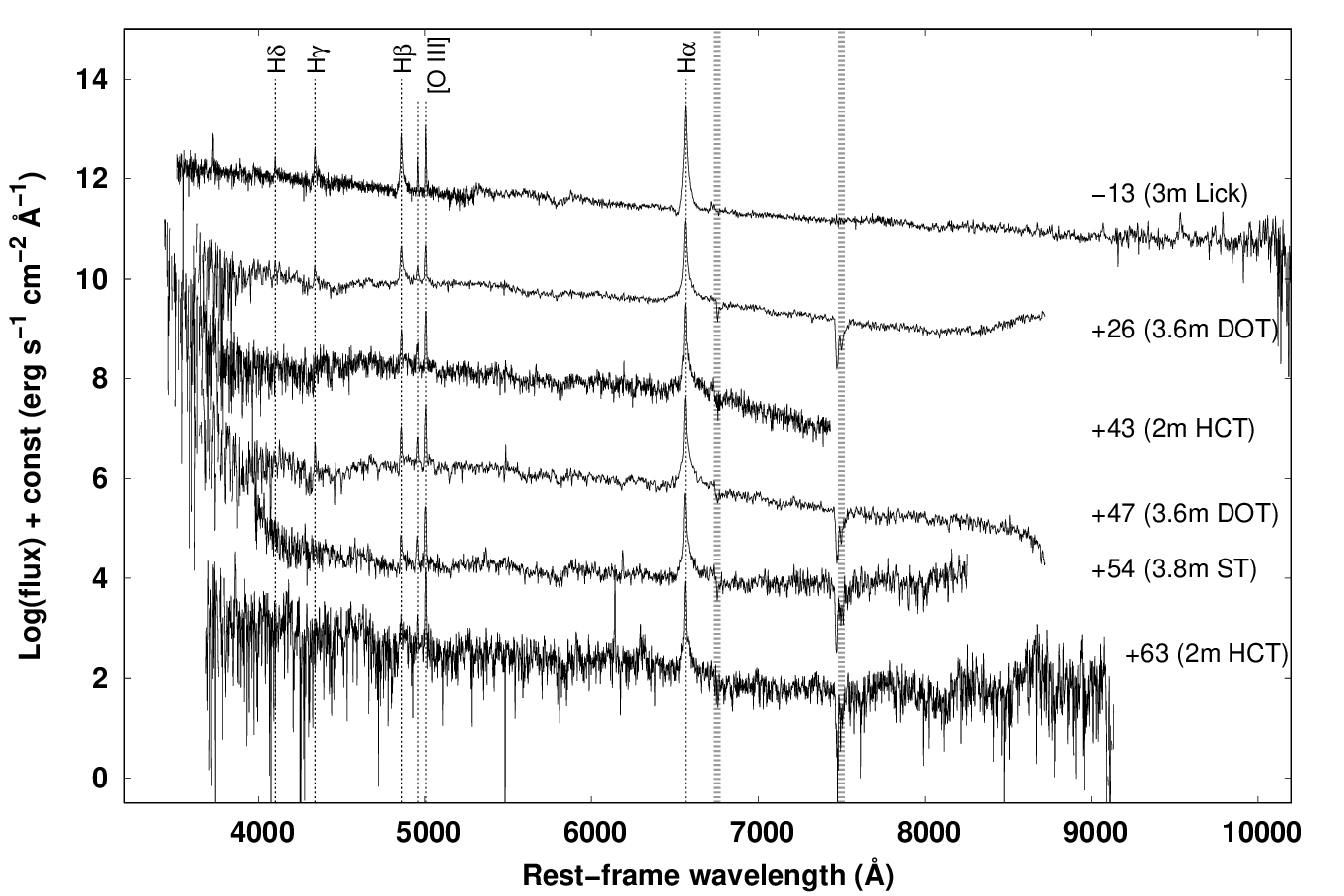}
    \caption{Spectroscopic evolution of SN 2023vbg. Each spectrum is labeled with the phase at which it was obtained and the telescope used for the observation. The thick gray lines indicate atmospheric absorption. The obtained spectra show no significant evolution, with narrow emission lines of the Balmer series and [O {\sc iii}] emission lines, likely originating from the host galaxy, present throughout all epochs. Also, $t=0$ d is SN peak.}
    \label{fig6}
\end{figure*}

\begin{figure}[htbp]
    \centering
    \includegraphics[width=0.5\textwidth]{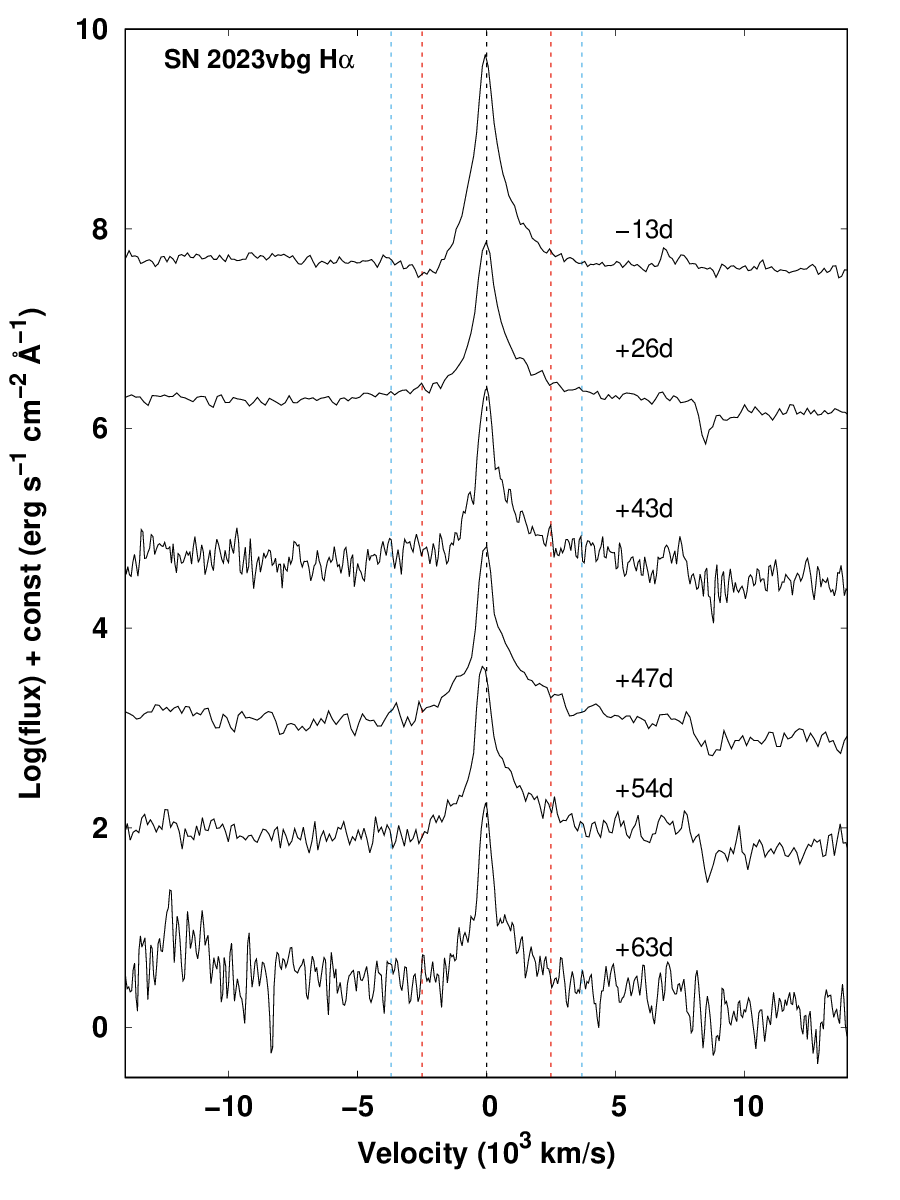}
    \caption{A zoomed-in view of the H$\alpha$ profile. The horizontal axis represents velocity, and the plot focuses on the region around H$\alpha$. The red dashed lines denote $\pm2500\ \mathrm{km\ s^{-1}}$, while the blue dashed lines denote $\pm3700\ \mathrm{km\ s^{-1}}$. Each epoch is given relative to the SN peak, as in Figure~\ref{fig6}.}
    \label{fig7}
\end{figure}

Figure~\ref{fig6} shows the spectral evolution of SN 2023vbg.
Prominent features are marked with lines for clarity.
Additionally, Figure~\ref{fig7} shows the zoomed-in profiles of H$\alpha$.
The first spectrum is the classification spectrum reported on TNS, which was taken during the `early-bump' phase.
Subsequent spectra were obtained during the decline phase.

Overall, the spectra show slow evolution, with narrow emission lines of the Balmer series visible throughout all epochs. The [O {\sc iii}] $\lambda\lambda4959,5007$ emission lines are likely of host galaxy origin.
In the first spectrum, a P-Cygni absorption feature is visible and the peak of this absorption is located at approximately $-2500\ \mathrm{km\ s^{-1}}$ (Figure~\ref{fig7}).
In the subsequent spectra, the P-Cygni absorption is no longer clearly visible, but the line profile is asymmetric with a suppressed blue wing.

\begin{table}[htpb]
    \begin{center}
        \caption{FWHM of H$\alpha$ line velocity }
        \label{tab:fwhm}
        \begin{tabular}{c | c c}
        \hline \hline
            Epoch & Narrow component & Intermediate component \\
            (days) & ($\mathrm{km\ s^{-1}}$) & ($\mathrm{km\ s^{-1}}$) \\
            \hline
            -13 & $<$370 & 1070 $\pm$31 \\
            +26 & $<$600 & 2430 $\pm$133 \\
            +43 & $<$500 & 2740 $\pm$112 \\
            +47 & $<$500 & 3140 $\pm$110 \\
            +54 & $<$500 & 3000 $\pm$126 \\
            +63 & $<$400 & 2310 $\pm$162 \\
            \hline
        \end{tabular}
    \end{center}
    \tablecomments{Each epoch is measured relative to the SN peak. Each epoch is measured relative to the SN peak. The intermediate-width component is fit with the Lorentzian profile in the earliest epoch ($\mathrm{t}=-13~\mathrm{d}$) and with the Gaussian profile in the other epochs.}
\end{table}

As shown in Figure \ref{fig7}, the H$\alpha$ line profile exhibits two components: a narrow component and an intermediate component. We performed the fitting to the line profile at each epoch using CURVE FIT in Python, with a combination of narrow (including the background host emission) and intermediate-width components (Table~\ref{tab:fwhm}). For the latter, we adopted either the Gaussian or Lorentzian profile; only for the earliest epoch ($\mathrm{t}=-13~\mathrm{d}$) the Lorentzian profile provides a better fit than the Gaussian profile, indicating a high density leading to multiple electron scatterings. The fit shows that the FWHM of the intermediate component in the post-maximum phase remains $< 3000 ~\mathrm{km s^{-1}}$ (but well resolved beyond the fitting uncertainty estimated with CURVE FIT) and did not show substantial evolution.

\subsection{Spectral Comparison}\label{ssec:spec compa}
\begin{figure*}[t]
    \centering
    \includegraphics[width=0.7\textwidth]{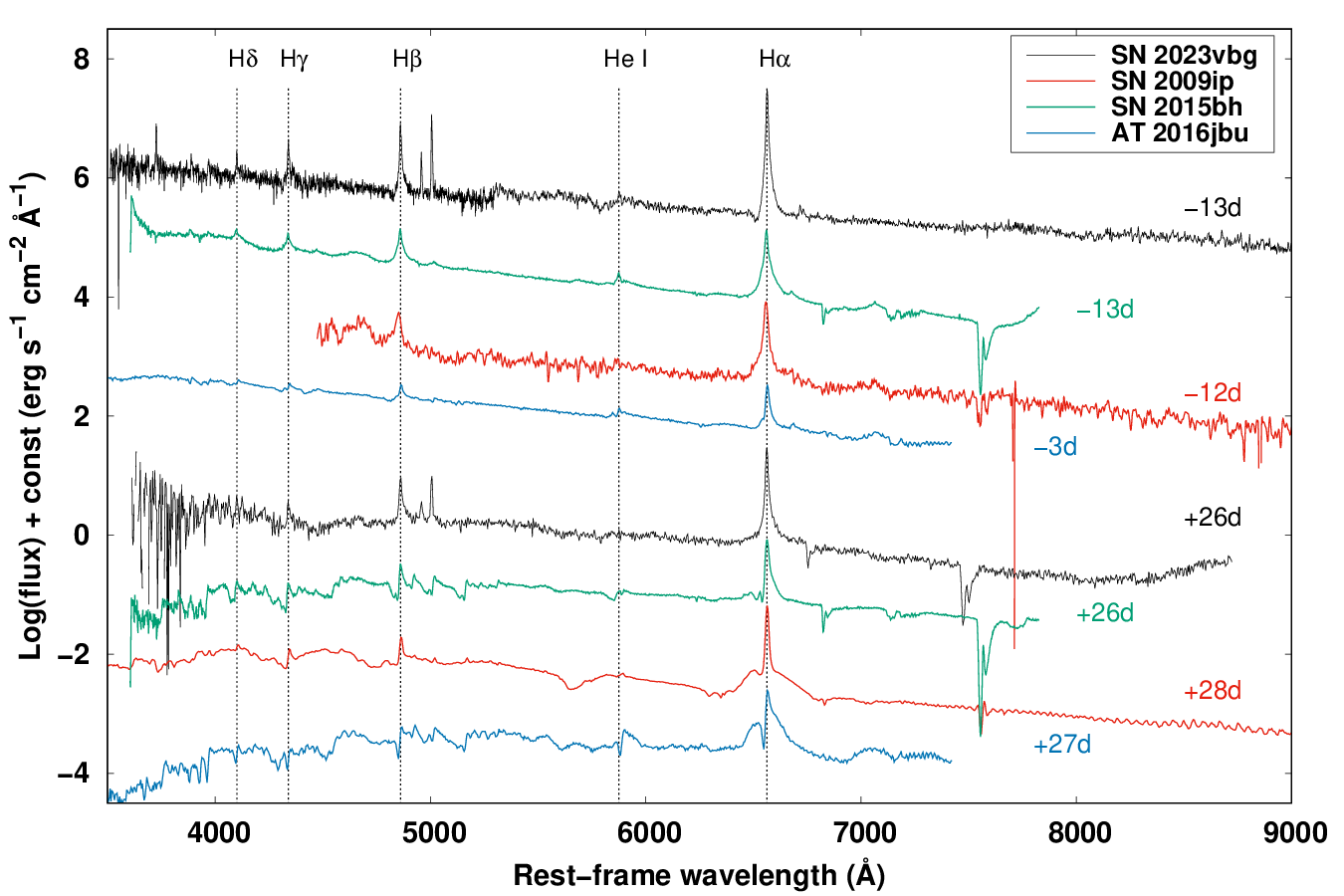}
    \includegraphics[width=0.7\textwidth]{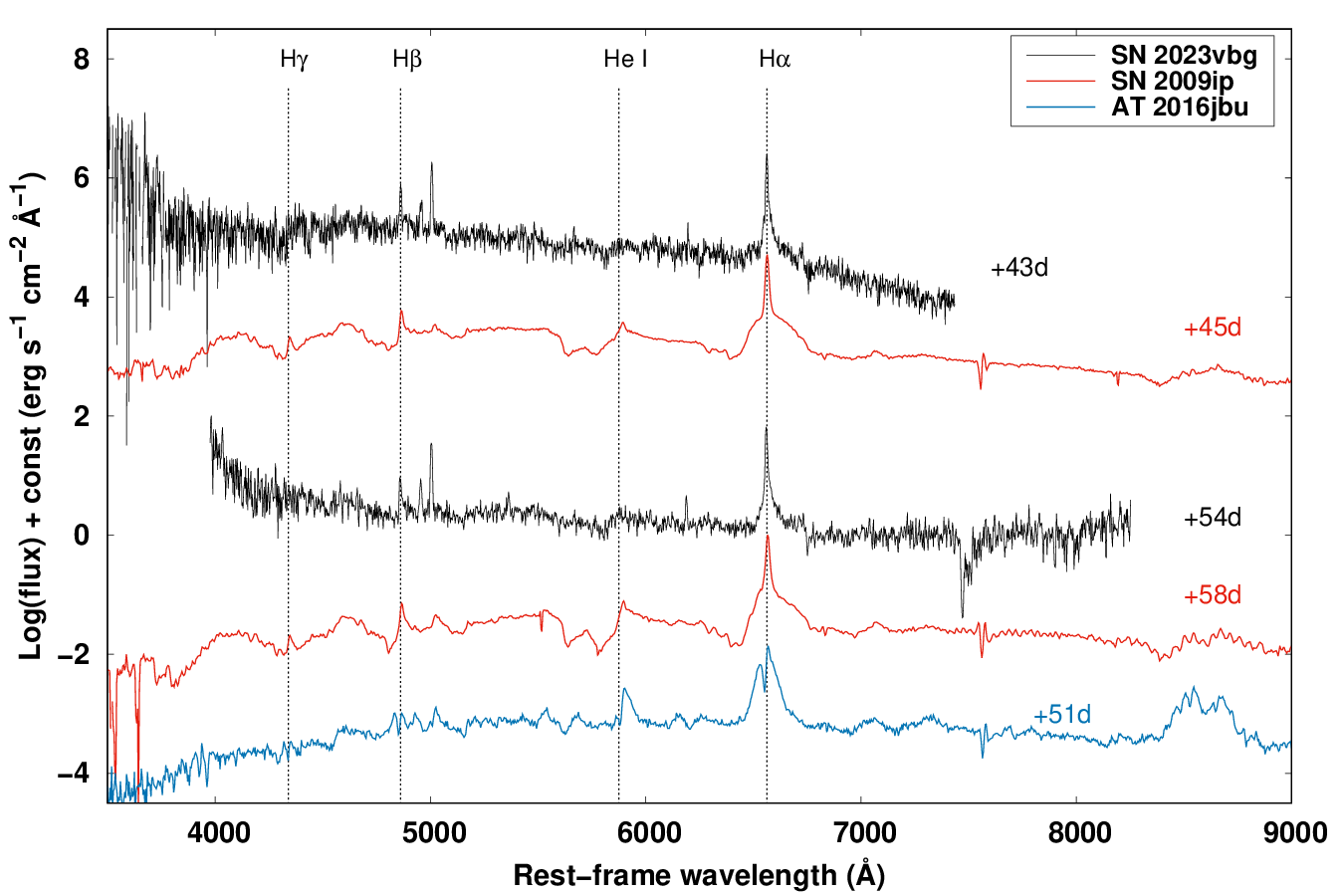}
    \caption{Upper: Comparison of early spectra of SN 2023vbg ($t=-13$ and $26$ d) with spectra of 09ip-like objects. The phase of each spectrum is labeled to the right, indicating the number of days since the respective peak as referenced in the literature. 
    Lower: Comparison of late spectra of SN 2023vbg ($t=43$ and $54$ d) with spectra of 09ip-like objects.}
    \label{fig8}
\end{figure*}

In Figure~\ref{fig8}, several epochs are selected for the comparison between spectra of SN 2023vbg and other 09ip-like objects, SN 2009ip \citep{mauerhan2013}, SN 2015bh \citep{thone2017}, AT 2016jbu \citep{brennan2022} at similar phases.
The top panel compares spectra from the pre-maximum phase to approximately $t=30$ d, and the bottom panel compares spectra from $t=30$ to $60$ d.
The earliest spectrum of SN 2023vbg ($t=-13$ d) displays a blue continuum and narrow hydrogen emission lines.

Additionally, a P-Cygni absorption feature in H$\alpha$ at around $-2500 ~\mathrm{km s^{-1}}$ is visible, along with marginal He~\textsc{i} $\lambda5876$ absorption. The absorption component seen in the rising phase just before the main peak is unique for SN 2023vbg among the 2009ip-like objects. On the other hand, some SN 2009ip-like objects show the P-Cygni profile in Ha in the pre-maximum `Event A' phase (i.e., earlier than for SN 2023vbg) frequently with much higher velocity; the examples include SNe 2009ip \citep[e.g.,][]{fraser2013} and 2023ldh \citep{pastorello2025}. The diversity might imply the presence or absence of materials along the line of sight, with different velocity for different objects. This, in turn, may point to an asymmetric CSM (such as a disk- or torus-like configuration) surrounding the progenitor and a part of the diversity might be associated with the viewing direction effect.

However, subsequent spectra of SN 2023vbg exhibited different evolution compared to the other objects.
In the post-maximum phases, the spectra of the comparison objects develop broad Balmer series components, along with blue-shifted (P-Cygni) absorption features (e.g., He {\sc i}, forest of Fe~{\sc ii}).
In contrast, SN 2023vbg primarily displayed only narrow and intermediate components of the Balmer series without any particularly prominent additional features.

\section{Discussion} \label{sec:disccusion}

\subsection{Mass-loss Rate} \label{ssec:mass-loss}
In previously observed 09ip-like objects, it is commonly believed that they underwent significant mass loss over a period of several years to months prior to the SN explosion \citep[e.g.,][]{margutti2014,elias-rosa2016,Fransson2022}. 
SN 2023vbg also exhibited a precursor with comparable luminosity, which, if attributed to the progenitor's mass loss, suggests that the progenitor of SN 2023vbg similarly underwent substantial mass loss prior to the explosion.

\citet{gangopadhyay2025} estimated the progenitor's mass-loss rate for SN 2021foa by assuming that the interaction between the CSM and ejecta is driven by the energy at the shock front.
Using similar assumptions, we estimate the mass-loss rate using the following equation \citep{chugai1994}:
\begin{equation}
\dot{M}=\frac{2L}{\epsilon}\frac{v_w}{v_{SN}^3}
\label{equ1}
\end{equation}
Here, $\dot{M}$ is the mass-loss rate, $\epsilon$ represents the efficiency of converting the shock's kinetic energy into radiative energy (assumed to be 0.5 in the present work). 
$v_w$ represents the wind velocity of the CSM, and we adopt a steady wind velocity typical of LBVs, $v_w\sim100\ \mathrm{km\ s^{-1}}$ \citep{vink2018}. 
$v_{\mathrm{SN}}$ is the velocity of SN ejecta; $v_{\mathrm{SN}}\sim 7000\ \mathrm{km\ s^{-1}}$ is adopted from the expansion velocity of the ejecta based on the BB fit. 
$L$ is the bolometric luminosity, and we use $L\sim1.1\times10^{43}\ \mathrm{erg\ s^{-1}}$ as estimated by integrating the SED at $t=3.8$\,d. 

We then obtain a mass-loss rate of $\dot{M}\sim0.02\ \mathrm{M_{\odot}\,yr^{-1}}$ corresponding to the CSM encountered by the shock at $t=3.8$\,d. This estimated value is in the same order as those previously estimated for SN 2009ip and other 09ip-like objects, e.g., $0.07\ \mathrm{M_{\odot}\,yr^{-1}}$ for SN 2009ip \citep{margutti2014}, and $0.04\ \mathrm{M_{\odot}\,yr^{-1}}$ for SN 2019zrk \citep{Fransson2022}.
Therefore, as discussed for these SNe, we conclude that SN 2023vbg also undergoes substantial mass loss prior to the explosion.

While a stellar evolution channel leading to such a high mass-loss rate seen for SNe IIn has not yet been clarified, comparison to some Galactic objects is useful. 
The derived mass-loss rate is significantly higher than the typical mass-loss rates of red supergiants and yellow hypergiants \citep[$10^{-4}$--$10^{-3}\ \mathrm{M_{\odot}\,yr^{-1}}$;][]{smith2014review}, as well as the wind during the quiescent phase of LBVs \citep[$10^{-5}$--$10^{-4}\ \mathrm{M_{\odot}\,yr^{-1}}$;][]{vink2018}. On the other hand, these values are of the same order as the mass-loss rates observed in giant eruptions of LBVs \citep[e.g.,][]{kiewe2012,smith2014review}. 

\subsection{The Origin of Early-bump} \label{ssec:early-bump}
As mentioned in Sect.~\ref{ssec:LC compa}, the remarkable `early-bump' is observed in SN 2023vbg before the SN peak. Non-smooth light curve evolution in the interacting SNe is frequently linked to (radially) non-smooth CSM distribution \citep[e.g.,][]{mauerhan2013}, since the CSM distribution is reflected to the light curve evolution through the SN-CSM interaction. We however note that discussion along this line has been done mainly to explain bump-like structures in the decline-phase light curves. 
The `early-bump' seen in SN 2023vbg is unique, since it may contain information on the CSM in the very vicinity of the SN progenitor; this makes the origin of the early-bump a particularly interesting question. 

\begin{figure}[htbp]
    \centering
    \includegraphics[width=0.5\textwidth]{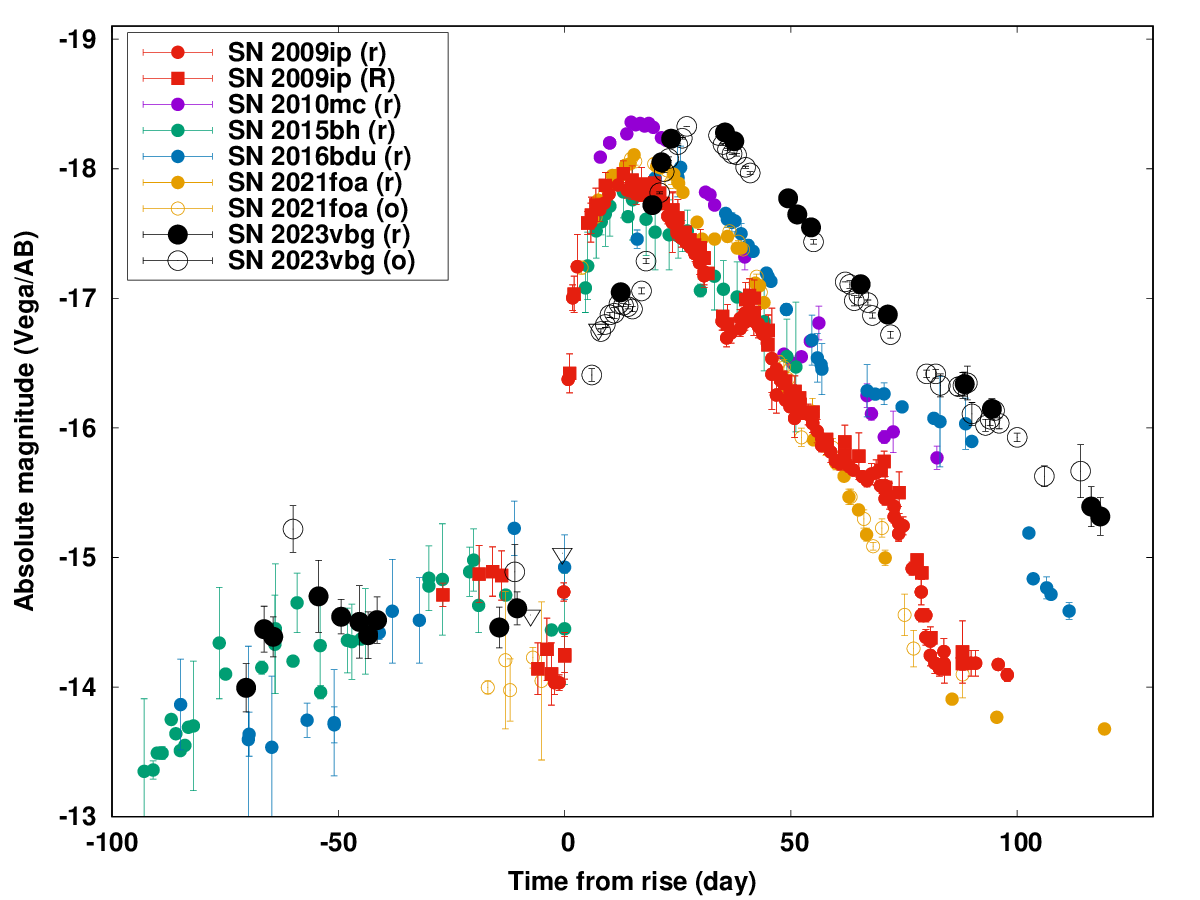}
    \caption{Comparing the $r$/$R$/$o$-band light curves of SN 2023vbg with those of other 09ip-like objects. Unlike in Figure~\ref{fig6}, $t=0$ here corresponds to the estimated explosion date for each object. For SN 2023vbg, we define $t=0$ at its last non-detection. Also, downward open triangles denote 3$\sigma$ upper limits.}
    \label{fig9}
\end{figure}

In the light curve analysis, it is physically intuitive to set the zero point of the light curve to the `explosion date', which is estimated by the extrapolation of the rising part to the SN peak back in time. For SN 2023vbg, we need an additional consideration -- the early bump might signal either the interaction between the materials ejected by two successive pre-SN outbursts, or the beginning of the SN-CSM interaction. Given its reaching to $\sim-17$ mag, the latter interpretation is more likely on the basis of energetics and the spectrum at the bump phase, and we proceed with this hypothesis. As compared to the range of the peak absolute magnitudes for Type IIP SNe \citep{anderson2014}, its magnitude in the early-bump is comparable to the brightest among SNe IIP (for which the contribution of the SN-CSM interaction is frequently introduced; \citep[e.g.,][]{jacobson2023,hiramatsu2023}). SN 2023vbg lies near the lower boundary of the r-band peak absolute-magnitude distribution for Type IIn supernovae discussed in \citet{hiramatsu2024}. These considerations further support that the SN-CSM interaction is the main power source in the bump phase. Then, the explosion date is estimated as the beginning of the rising behavior toward the early bump; $-27~\mathrm{d}$, from the rapid rise observed by ATLAS (see Figure 1).

Figure~\ref{fig9} compares the $r$/$R$/$o$-band light curve of SN 2023vbg with those of other 09ip-like objects, but with $t=0$ shifted to the estimated SN explosion date. This way, a larger degree of the diversity, highlighted by SN 2023vbg, is inferred than seen in Figure \ref{fig3}. Measured from the estimated explosion date, SN 2023vbg indeed showed a longer timescale to the SN peak than the other objects; this indicates that the column density of the CSM as measured outward from the position of the wind breakout \citep[e.g.,][]{moriya2014} might be larger than the other SNe. At the end of the early bump of SN 2023vbg, the rate of the luminosity rise was accelerated; this indicates that the diffusion timescale within the unshocked CSM showed a sudden decrease. Another hint is on the mass-loss rate measured around the SN peak (or simply the peak magnitude), which is comparable among the 09ip-like objects including SN 2023vbg. 

From the above considerations, the following picture has emerged. The larger column density of the CSM for SN 2023vbg despite the similar CSM density at the relatively outer region (corresponding to the peak) with the other 09ip-like objects (Section \ref{ssec:mass-loss}) indicates that the CSM distribution around SN 2023vbg may be described by two components \citep[e.g.,][]{moriya2015}. In the inner part, there is a high-density CSM component which is denser than those of other 09ip-like objects. The transition of the light curve behavior (i.e., with accelerated rising) at $\sim13$ days after the estimated explosion date indicates that this inner dense CSM component extended up to $\sim8 \times 10^{14}$ cm, assuming that this transition took place when the inner CSM was swept up by the shock wave (with $v_{\rm SN} \sim7000$~km s$^{-1}$), leading to a sudden decrease of the optical depth and diffusion timescale within the unshocked CSM. 

Following this interpretation, a key difference of SN 2023vbg from other 09ip-like objects is its having the inner dense CSM component, and such a component is not clearly seen in the others despite their sharing the pre-SN activity. We suggest that the others may also have such a dense, inner CSM component, but it is located even closer to the progenitor star ($\ll 8 \times 10^{14}$ cm) than in the case of SN 2023vbg. In this case, the inner CSM component might have been swept up already before the wind breakout, leaving no observational trace of the inner component.

\subsection{Nature of SN 2023vbg} \label{ssec:nature of 2023vbg}
The above picture on the difference between SN 2023vbg and other 09ip-like objects (Section \ref{ssec:early-bump}) is consistent with the properties of their precursors. SN 2023vbg shows a hint of a higher level of the precursor emission starting earlier than observed in other objects (e.g., Figure \ref{fig2}). Assuming that this is related to the eruption of materials \citep[e.g.,][]{smith2014review} and it had lasted for $\sim1000$ days before SN 2023vbg (Section \ref{ssec:precursor}), this translates into the physical scale of $\sim8 \times 10^{14}$ cm (with $v_{\rm w}=100$~km s$^{-1}$) that matches the estimated extension of the inner CSM component for SN 2023vbg. 

If the possible longer timescale of the pre-SN activity in SN 2023vbg than in the comparison objects would apply generally in the final evolution of the progenitor, the outer CSM component may also be more extended than in the other objects. This might explain the smooth evolution in the decay phase of the light curve without a rapid decay found in the comparison objects at $\sim50-100$ days after the SN peak; we might see the outer edge of the outer CSM component swept up for the comparison objects, while the situation is not yet reached for SN 2023vbg. This might further be linked to the absence of broad spectral features in SN 2023vbg even in the late phase -- due to the extended outer CSM, the photosphere may still be located within the CSM. 

Additionally, we do not see clear variability in the post-peak light curve of SN 2023vbg, unlike some of the comparison objects (Section \ref{ssec:LC compa}). This difference is likely attributed to variations in the density structure of the outer CSM \citep[e.g.,][]{mauerhan2013,graham2014}. The relative smooth light curve of SN 2023vbg suggests that the CSM regions interacting with the ejecta during the observed period did not exhibit significant density inhomogeneities. In terms of the pre-SN evolution, this indicates that the pre-SN activity might have been generally less variable for SN 2023vbg than the others. A hint on this is seen in Figure \ref{fig2} and the related discussion in Section \ref{ssec:precursor}; if the long-term ($>1000$ days) precursor is real, it indicates a lower level of variability than in SN 2009ip. 

In summary, we suggest that some observational characteristics of SN 2023vbg that differentiate this object from other 09ip-like objects may be explained by its pre-SN activities being at a higher level, more extended in timescale, and less variable than the others. This highlights the potentially diverse nature of the progenitor channels toward SN09-ip like objects (e.g., progenitor mass, binary); while investigating the origin of these differences in the pre-SN activity is beyond our scope here, the present work provides a potentially important data set for further investigation of this still-unclarified issue in stellar evolution. 

\section{Summary and Conclusion} \label{sec:summary and conclusion}
From photometric and spectroscopic observations, we have found that SN 2023vbg is a 09ip-like SN. It shares many properties with other 09ip-like objects, but importantly, it also shows some striking differences, pointing to diverse properties of the class. 
The detection of a precursor and the overall timescale of the light curve were very similar.
However, the presence of an `early-bump' of the LC rising phase was unique compared with other 09ip-like objects.
Additionally, unlike the comparison objects, the spectra exhibited slow evolution, and no broad velocity components were seen.

Based on the analyses of these observational properties, we have presented a scenario for SN 2023vbg. The overall similarity to other 09ip-like objects indicates a common progenitor channel, e.g., a massive star (e.g., LBV) that experienced substantial mass loss (at least) several years before the explosion. The characteristic observational properties of SN 2023vbg could be explained by its double-structured CSM, as well as otherwise generally smooth CSM density distribution in a more extended scale. The double structure (inner and outer CSM components) may indeed be shared by other objects, but a more confined nature of the inner component in the other objects might not allow detection of clear signatures associated with such a component. These key differences should be translated into the difference in the pre-SN activities, highlighting the diverse nature of the progenitor channel toward 09ip-like objects. 


The nature of 09ip-like SNe remains poorly understood. Further observational effort, together with modeling observables and constructing progenitor evolution models, will be necessary for better understanding these interesting objects. In this aspect, the Vera C. Rubin Observatory’s Legacy Survey of Space and Time; \cite{ivezi2019} will change the game -- both the precursors and post-peak high-quality light curves will be automatically obtained for nearby 09ip-like objects like SN 2023vbg. As the number of such objects grows and more detailed observations become available from the precursor stage onward, this will be crucial for advancing our understanding of progenitor evolution prior to the SN explosion and CSM formation.

\section{acknowledgement}
 We are grateful to graduate and undergraduate students of Kagoshima University (KU) for performing  the optical and NIR observations. This work was partly supported by Grant-in-Aid for Scientific Research (C) 22K03676. KM acknowledges support from the Japan Society for the Promotion of Science (JSPS) KAKENHI grant Nos. 24KK0070 and 24H01810. The KU 1-m and Seimei telescope are members of the Optical and Infrared Synergetic Telescopes for Education and Research (OISTER) program funded by the MEXT of Japan. The ZTF forced-photometry service was funded under the Heising-Simons Foundation grant 12540303 (PI: Graham). D.K.S. acknowledges the support provided by DST-JSPS under grant number DST/INT/JSPS/P363/2022. This work was supported by JSPS Bilateral Program Number JPJSBP 120227709. The spectra taken by the Seimei telescope are obtained with the KASTOR
(Kanata And Seimei Transient Observation Regime) campaign (24A-N-CT15,
PI: K.M.).

\bibliographystyle{aasjournal}
\bibliography{sn2023vbg}

\appendix
\section{observational data}

\begin{table*}[ht]
    \centering
    \caption{Log of spectroscopic observations of SN 2023vbg}
    \label{tab:speclog}
    \centering
    \begin{tabular}{c c c c c c}
        \hline \hline
        Date & MJD & Phase & Telescope + instrument & Spectral Range & Exposure Time \\
         & & (d) & & (\AA) & (s) \\
        \hline
        2024 Feb 16 & 60357.3 & +26 & DOT + ADFOSC & 3500-7100 & 1200 \\
        2024 Mar  4 & 60374.11 & +43 & HCT + HFOSC & 3800-6840 & 3600 \\
        2024 Mar  8 & 60378.23 & +47 & DOT + ADFOSC & 3500-7100 & 900$\times$2 \\
        2024 Mar 15 & 60384.5 & +54 & Seimei + KOOLS-IFU & 4100-8900 & 600 \\
        2024 Mar 24 & 60394.43 & +63 & HCT + HFOSC & 3800-6840 & 2700 \\
        \hline
    \end{tabular}
\end{table*}

\begin{table*}[ht]
    \centering
    \caption{ARIES photometry of SN 2023vbg}
    \label{tab:ARIES photometry list}
    \centering
    \begin{tabular}{c c c c c c}
        \hline \hline
        Date & MJD & Phase & U & B & V \\
         & & (d) & (mag) & (mag) & (mag) \\
        \hline
        2024 Feb 7 & 60347.83 & +16.6 & 16.592 \ $\pm$ \ 0.041 & 16.813 \ $\pm$ \ 0.028 & 16.385 \ $\pm$ \ 0.014 \\
        2024 Feb 16 & 60356.82 & +25.6 & - & 17.219 \ $\pm$ \ 0.028 & 16.769 \ $\pm$ \ 0.015 \\
        \hline
    \end{tabular}
\end{table*}

\begin{table}[ht]
    \centering
    \caption{Iriki photometry of SN 2023vbg}
    \label{tab:iriki photometry list}
    \centering
    \begin{tabular}{ccccccc}
        \hline \hline
        MJD & Phase & g & i & J & H & Ks \\
         & (d) & (mag) & (mag) & (mag) & (mag) & (mag) \\
         \hline
        60322.57 & -7.8  & - & - & 16.079 \ $\pm$ \ 0.020 & 15.799 \ $\pm$ \ 0.020 & 15.626 \ $\pm$ \ 0.030 \\
        60326.48 & -3.9  & - & - & 15.706 \ $\pm$ \ 0.016 & 15.567 \ $\pm$ \ 0.018 & 15.299 \ $\pm$ \ 0.023 \\
        60334.60 &  4.2  & 15.992 \ $\pm$ \ 0.009 & 16.139 \ $\pm$ \ 0.014 & 15.566 \ $\pm$ \ 0.017 & 15.425 \ $\pm$ \ 0.019 & 15.140 \ $\pm$ \ 0.014 \\
        60335.64 &  5.3  & 16.028 \ $\pm$ \ 0.006 & 16.142 \ $\pm$ \ 0.007 & 15.528 \ $\pm$ \ 0.016 & 15.448 \ $\pm$ \ 0.019 & 15.221 \ $\pm$ \ 0.018 \\
        60337.64 &  7.3  & 16.113 \ $\pm$ \ 0.008 & 16.192 \ $\pm$ \ 0.010 & - & 15.472 \ $\pm$ \ 0.019 & 15.276 \ $\pm$ \ 0.017 \\
        60347.60 & 17.2  & 16.557 \ $\pm$ \ 0.016 & 16.515 \ $\pm$ \ 0.015 & 15.809 \ $\pm$ \ 0.018 & 15.726 \ $\pm$ \ 0.025 & 15.439 \ $\pm$ \ 0.029 \\
        60348.63 & 18.3  & 16.609 \ $\pm$ \ 0.009 & 16.536 \ $\pm$ \ 0.010 & 15.854 \ $\pm$ \ 0.017 & 15.737 \ $\pm$ \ 0.019 & 15.519 \ $\pm$ \ 0.027 \\
        60352.67 & 22.3  & 16.772 \ $\pm$ \ 0.008 & 16.675 \ $\pm$ \ 0.009 & 15.948 \ $\pm$ \ 0.018 & 15.890 \ $\pm$ \ 0.021 & 15.630 \ $\pm$ \ 0.021 \\
        60356.55 & 26.2  & - & - & 16.081 \ $\pm$ \ 0.013 & 15.980 \ $\pm$ \ 0.027 & 15.749 \ $\pm$ \ 0.035 \\
        \hline
    \end{tabular}
\end{table}

\begin{table}[ht]
    \centering
    \caption{Swift photometry of SN 2023vbg}
    \label{tab:UVOT photometry list}
    \begin{tabular}{cccccccc}
        \hline \hline
        MJD & Phase & uvw2 & uvm2 & uvw1 & U & B & V \\
         & (d) & (mag) & (mag) & (mag) & (mag) & (mag) & (mag) \\
        \hline
        60322.2 & -8.2  & 15.46 \ $\pm$ \ 0.05 & 15.46 \ $\pm$ \ 0.09 & 15.50 \ $\pm$ \ 0.05 & 15.76 \ $\pm$ \ 0.05 & 16.96 \ $\pm$ \ 0.07 & 16.89 \ $\pm$ \ 0.12 \\
        60326.1 & -4.3  & 15.43 \ $\pm$ \ 0.06 & 15.13 \ $\pm$ \ 0.09 & 15.14 \ $\pm$ \ 0.06 & 15.26 \ $\pm$ \ 0.05 & 16.41 \ $\pm$ \ 0.07 & 16.41 \ $\pm$ \ 0.13 \\
        60328.5 & -1.9  & 15.25 \ $\pm$ \ 0.05 & - & - & - & 16.26 \ $\pm$ \ 0.07 & 16.18 \ $\pm$ \ 0.11 \\
        60335.4 & 5.0   & 15.28 \ $\pm$ \ 0.06 & 15.86 \ $\pm$ \ 0.09 & 15.42 \ $\pm$ \ 0.06 & 15.33 \ $\pm$ \ 0.05 & 16.34 \ $\pm$ \ 0.06 & 16.25 \ $\pm$ \ 0.10 \\
        60339.9 & 9.5   & 16.32 \ $\pm$ \ 0.08 & 16.08 \ $\pm$ \ 0.09 & 15.89 \ $\pm$ \ 0.07 & - & - & - \\
        60342.1 & 11.7  & 16.58 \ $\pm$ \ 0.08 & 16.67 \ $\pm$ \ 0.09 & 16.25 \ $\pm$ \ 0.07 & - & - & - \\
        60346.3 & 15.7  & 17.00 \ $\pm$ \ 0.09 & 17.00 \ $\pm$ \ 0.10 & 16.51 \ $\pm$ \ 0.07 & - & - & - \\
        60348.1 & 17.6  & 17.39 \ $\pm$ \ 0.12 & 16.73 \ $\pm$ \ 0.10 & 17.52 \ $\pm$ \ 0.14 & - & - & - \\
        60350.8 & 20.3  & - & 17.46 \ $\pm$ \ 0.10 & - & - & - & - \\
        60354.1 & 23.7  & 17.53 \ $\pm$ \ 0.09 & 17.50 \ $\pm$ \ 0.09 & 17.14 \ $\pm$ \ 0.07 & - & - & - \\
        60363.9 & 33.5  & 18.31 \ $\pm$ \ 0.19 & 18.03 \ $\pm$ \ 0.14 & 17.71 \ $\pm$ \ 0.17 & 17.41 \ $\pm$ \ 0.17 & 17.71 \ $\pm$ \ 0.15 & - \\
        60366.8 & 36.4  & - & 17.96 \ $\pm$ \ 0.18 & - & - & 17.41 \ $\pm$ \ 0.16 & - \\
        \hline
    \end{tabular}
\end{table}

\end{document}